\magnification=1200
\def\tr#1{{\rm tr} #1}
\def\d{\partial}
\def\f#1#2{{\textstyle{#1\over #2}}}

\def\next{\hfil\break\noindent}
\def\R{{\bf R}}
\def\Z{{\bf Z}}
\def\Quadrat#1#2{{\vcenter{\hrule height #2
  \hbox{\vrule width #2 height #1 \kern#1
    \vrule width #2}
  \hrule height #2}}}
\def\halmos{\ \ $\Quadrat{4pt}{0.4pt}$}
\font\title=cmbx12

{\title 
\centerline{Existence of constant mean curvature foliations in spacetimes}
\centerline{with two-dimensional local symmetry }}

\vskip 1cm

\noindent
Alan D. Rendall
\next
Institut des Hautes Etudes Scientifiques
\next
35 Route de Chartres
\next
91440 Bures sur Yvette
\next
France

\vskip 10pt\noindent
and

\vskip 10pt\noindent
Max-Planck-Institut f\"ur Gravitationsphysik
\next
Schlaatzweg 1
\next
14473 Potsdam
\next
Germany\footnote{*}{Present address}

\vskip 1cm\noindent
{\bf Abstract} 

\noindent
It is shown that in a class of maximal globally hyperbolic
spacetimes admitting two local Killing vectors, the past (defined with
respect to an appropriate time orientation) of any compact constant mean 
curvature hypersurface can be covered by a foliation of compact constant 
mean curvature hypersurfaces. Moreover, the mean curvature of the leaves of
this foliation takes on arbitrarily negative values and so the initial 
singularity in these spacetimes is a crushing singularity. The simplest
examples occur when the spatial topology is that of a torus, with the
standard global Killing vectors, but more exotic topologies are also
covered. In the course of the proof it is shown that in this class of
spacetimes a kind of positive mass theorem holds. The symmetry singles out 
a compact surface passing through any given point of spacetime and the
Hawking mass of any such surface is non-negative. If the Hawking mass of
any one of these surfaces is zero then the entire spacetime is flat. 

\vskip 1cm\noindent
{\bf 1. Introduction} 

There are a number of general results in the literature on the
properties of foliations by compact spacelike hypersurfaces of
constant mean curvature (CMC hypersurfaces) in spacetimes which admit
a compact Cauchy hypersurface.  (See [16], [1] and references
therein.) However, the only results which give criteria in terms of Cauchy 
data for the existence of such foliations covering more than a small 
neighbourhood of a given CMC hypersurface are restricted to
special classes of spacetimes, all of which have high symmetry. The
results of this paper also apply only to certain spacetimes with
symmetry but represent a significant generalization, since they
include for the first time spacetimes containing both matter and
gravitational waves. The method used suggests that there is a
close connection between the question of global existence of CMC
foliations and that of global existence of solutions of the
Einstein-matter equations and such a connection helps to explain why
it has up to now been necessary to make symmetry assumptions: we
cannot understand the question of global existence of CMC foliations
in a context where we do not understand the question of global
existence for the Einstein-matter equations. When a CMC foliation
exists in a spatially compact spacetime satisfying the strong energy
condition it is unique, provided the exceptional case of flat
spacetime is excluded. It provides an invariantly defined preferred
time coordinate on spacetime.

The spacetimes studied in the following are defined by two conditions. The
first is that they be solutions of the Einstein equations coupled to certain 
matter fields and the second is that they admit a compact CMC Cauchy 
hypersurface which possesses a two-dimensional abelian group of local 
symmetries without fixed points. (The second condition, stated here 
informally, is made precise in the next section.) The simplest example 
of a symmetry of this kind is that where the compact  Cauchy hypersurface is 
diffeomorphic to the torus $T^3=S^1\times S^1\times S^1$, with the symmetries 
given by the action of $U(1)\times U(1)$ acting on two of the three $S^1$ 
factors by rotations. As shown below, the mean curvature of the Cauchy 
hypersurface must be non-zero, except in the trivial case where the spacetime 
is flat. Without loss of generality, reversing the time orientation if 
necessary, it can be assumed to be negative. The main result of this paper 
(Theorems 5.1 and 6.1) is that if the spacetime is the maximal globally 
hyperbolic development of data with local $U(1)\times U(1)$ symmetry on a CMC 
Cauchy hypersurface then the entire past of this hypersurface can be covered 
by a CMC foliation, with the mean curvature taking all values in the interval 
$(-\infty,H_0]$, where $H_0$ is the mean curvature of the initial 
hypersurface. In particular the initial singularity in these spacetimes is a 
crushing singularity in the sense of Eardley and Smarr[7]. The assumption
made on the matter model is that it is either collisionless matter modelled
by the Vlasov equation (Theorem 5.1) or a wave map with values in an
arbitrary complete Riemannian manifold (Theorem 6.1). It is also shown that
the CMC foliation can be extended so that the mean curvature takes on all
values in the interval $(-\infty,0)$. Unfortunately this does not by itself
suffice to show that the CMC foliation covers the entire future of the
initial hypersurface.

Special cases of this result are already known. The first is that of the
Gowdy spacetimes on the torus. These are vacuum spacetimes with global 
$U(1)\times U(1)$ symmetry which satisfy the additional condition that the
so-called twist constants vanish (see below). The result was proved in this
case by Isenberg and Moncrief [12]. The second is that of solutions of the
Einstein-Vlasov system with plane symmetry [17]. In the first of these cases
there is no matter present, while in the second there are no gravitational 
waves. The results of this paper contain both these results as special
cases. It should be noted that they go beyond previous results even in
the vacuum case in two ways: they require only local, rather than global 
symmetry and they allow non-vanishing twist constants. The essential new 
element in comparison with the cases considered previously is the occurrence 
of nonlinear hyperbolic equations which are coupled to the matter equations.
These are treated with the help of methods introduced by Gu [11] in the study 
of wave maps and by Glassey and Strauss [10] in the study of the Vlasov-Maxwell
system.

As a by-product of this analysis, a theorem on the positivity of the
Hawking mass in spacetimes with local $U(1)\times U(1)$ symmetry is obtained. 
(Proposition 3.1). This says that the Hawking mass of any surface of symmetry 
in a spacetime of this type is non-negative and that if the Hawking mass 
vanishes for any one of these surfaces in a spacetime, then the spacetime is 
flat.

\vskip 10pt\noindent
{\bf 2. Local $U(1)\times U(1)$ symmetry}

The spacetimes considered in the following are defined on manifolds
of the form $M=\R\times S$, where $S$ is a bundle over the circle $S^1$
whose fibre is a compact orientable surface $F$. Let $p$ be the
projection of the universal cover $\tilde F$ onto $F$. Let $g_{\alpha\beta}$
be a globally hyperbolic metric on $M$ for which each submanifold 
$\{t\}\times S$ is a Cauchy hypersurface. $S$ is covered by its pull-back to a
bundle over $\R$. Since $\R$ is contractible, the latter bundle is isomorphic
to $\R\times F$. This is turn is covered in a natural way by $\R\times\tilde
F$, which is simply connected. Hence the universal cover $\tilde S$ of $S$
can be identified with $\R\times \tilde F$ and there is a natural fibre
preserving projection corresponding to $p$. Let $\hat p$ be the associated
projection of $\tilde M=\R^2\times\tilde F$ onto $M$. Define 
$\hat g_{\alpha\beta}$ to be the pull-back of $g_{\alpha\beta}$ by $\hat p$. 
Suppose that a two-dimensional Lie group $G$ acts effectively on $\tilde M$ 
by isometries of $\hat g_{\alpha\beta}$ in such a way that the orbits
are the inverse images under $\hat p$ of the fibres of the bundle
$S$. Each orbit with its induced metric is a simply connected 
Riemannian manifold of constant curvature and thus must be, up to
a constant conformal rescaling, isometric to the standard metric
on the sphere, the Euclidean plane or the hyperbolic plane. The
isometry group of the sphere has no two-dimensional subgroups and
thus the case $F=S^2$ is not possible. (If $G$ is replaced by a 
three-dimensional Lie group then $F=S^2$ is possible and the spherically 
symmetric spacetimes studied in [17] are obtained.) The surfaces
diffeomorphic to $F$ which correspond to the fibres of $S$ and whose
inverse images are the group orbits will be referred to as surfaces
of symmetry. These spacetimes will be said to have two-dimensional
local symmetry. In the case where the orbits are isometric to the
Euclidean plane, they will be said to have local $U(1)\times U(1)$
symmetry.

Consider now any spacelike hypersurface $S$ in the spacetime
$(M,g_{\alpha\beta})$ which is a union of surfaces of symmetry.
Choose one of these surfaces of symmetry and call it $F_0$.  Let
$\gamma$ be an affinely parametrized geodesic of the induced metric on
$S$ which starts orthogonal to $F_0$. It is also orthogonal to all the
other surfaces of symmetry which it meets. Taking all geodesics of
this type and following them until they meet $F_0$ again gives a
smooth mapping $\phi$ from $F_0\times I$ to $S$, where $I$ is some
interval. Let the parameter along the geodesics be chosen so that
$I=[0,2\pi]$. Let $\psi$ denote the mapping which takes $\phi(x,0)$ to
$\phi(x,2\pi)$ and let $\tilde\psi$ denote the lift of this mapping to a
diffeomorphism between two inverse images of $F_0$ in $\tilde M$ defined
by following geodesics in the covering space.  The
mapping $\tilde\psi$ maps the Killing vectors of $F_0$ corresponding
to the group action on one of the inverse images of $F_0$ in $\tilde
M$ to those corresponding to another inverse image. These two inverse
images can be identified with each other by an isometry, which is uniquely
determined up to an element of the isometry group of the Euclidean or 
hyperbolic plane respectively by their induced metrics. In the case that 
$\tilde F_0$ is isometric to the Euclidean plane $\tilde\psi_0$ is the 
composition of a linear mapping $\tilde\psi_1$ with a translation 
$\tilde\psi_2$. The mapping $\tilde\psi$ must preserve the
lattice of inverse images in $\tilde F_0$ of a given point in $F_0$.
These lattices are isometric, and so by using the freedom in identifying
the two covering spaces, it can be assumed without loss of generality that
they are identified with each other. It follows that $\tilde\psi_1$ can be 
represented as an element of $GL(2,\Z)$. When this element is not the 
identity it means in general that the topology of $S$ is that of a non-trivial
torus bundle over the circle. Let $y^A$ be periodic coordinates on $F_0$ 
corresponding to Cartesian coordinates on $\tilde F_0$. Let $(x,y^A)$ be Gauss
coordinates based on $F_0$ such that $y^A$ restrict to the
previously chosen coordinates on $F_0$. Now let $B^4=\det (g_{AB})$,
where upper case Roman indices take the values $2$ and $3$. The metric
takes the form:
$$dx^2+B^2\tilde g_{AB}(x)dy^Ady^B\eqno(2.1)$$
where $\det\tilde g_{AB}=1$. Let $L$ be the length of a geodesic which
starts normal to an orbit and ends when it intersects the same orbit 
again. Define $a=2\pi/(\int_0^L B^{-1}(x) dx)$ and:
$$x'=a\int_0^x B^{-1}(y) dy\eqno(2.2)$$ 
If $x'$ is used as a coordinate and the primes omitted from the notation
the metric takes the form:
$$A^2(dx^2+a^2\tilde g_{AB}dy^Ady^B)\eqno(2.3)$$
where $A$ is $a^{-1}B$, reexpressed as a function of $x'$. The new coordinate 
$x$ runs from $0$ to $2\pi$. The functions $\tilde g_{AB}(x)$
satisfy the relation $\tilde g_{AB}(x+2\pi)=n_A^Cn_B^D\tilde g_{CD}(x)$, where
$n^A_B$ are the components of a matrix in $GL(2,\Z)$. $A$ satisfies 
$A(x+2\pi)=A(x)$. A similar analysis for the case that $\tilde F_0$ is the 
hyperbolic plane would no doubt be more complicated and is not attempted 
here. However it apppears that, due to the fact that vector fields and
tracefree symmetric rank two tensors on a surface of genus higher than one
must have zeroes, in that case nothing will be obtained which goes beyond
the spacetimes with hyperbolic symmetry already studied in [17].  

The metric $\tilde g_{AB}$ can be parametrized in terms of two
functions $W$ and $V$ in the following way:
$$\tilde g_{22}=e^W\cosh V,\ \ \ \tilde g_{33}=e^{-W}\cosh V,\ \ \
\tilde g_{23}=\sinh V\eqno(2.4)$$
The values of $V$ and $W$ at $x=0$ and $x=2\pi$ are related by a 
diffeomorphism $N$ which does not have a simple explicit form. There
are several special cases which are of interest. Consider first the
case where the matrix $N$ with components $n^A_B$ is the identity. Then there 
is a natural action of $U(1)\times U(1)$ on the spacetime and we have the 
case of (global) $U(1)\times U(1)$ symmetry. A further specialization is 
given by the assumption that the reflections $y^A\mapsto -y^A$ are isometries
of the spacetime metric for $A=2,3$. Spacetimes satisfying this condition
will be called polarized $U(1)\times U(1)$-symmetric spacetimes. They
have the property that $V=0$. The vacuum spacetimes of
this class are the polarized Gowdy spacetimes[6]. The plane symmetric 
spacetimes studied in [17] have the property that $W=V=0$. 
When $N$ is not the identity, there are two qualitatively
different cases. If $N$ is diagonalizable and not the identity,
then either it is minus the identity, or the two eigenvalues are
distinct. If it is minus the identity then $S$ has a two-fold cover
which is a torus and for our purposes is essentially the same as
when $N$ is the identity. When the eigenvalues are distinct
the manifold $S$ admits a geometric structure of type Sol in the sense of 
Thurston [18]. The metrics obtained in that case include ones
which are of Bianchi type VI${}_0$. There is a polarized case, where
reflections in the eigendirections of $N$ are supposed to be 
isometries of the metric $\hat g_{\alpha\beta}$ on the universal
covering space. If $N$ has a non-standard Jordan form then, by
passing to a two-fold cover if necessary, we can assume that these
eigenvalues are equal to unity. The resulting manifold $S$ admits
a geometric structure of type Nil [18]. The metrics obtained in
that case include those of Bianchi type II.

\vskip 10pt\noindent
{\bf Lemma 2.1} {\it Let $(M,g)$ be a non-flat spacetime with local
$U(1)\times U(1)$ symmetry having a symmetric constant mean curvature Cauchy 
hypersurface and satisfying the dominant and strong energy conditions. Then 
given any point $p$ on the Cauchy surface there exists an open neighbourhood 
$U$ of $p$ and a smooth local diffeomorphism $\phi$ of 
$I\times [0,2\pi]\times T^2$ onto $U$ for some interval $I$ such that:

\noindent
(i) if $\phi(t,x_1,y_1)=\phi(t,x_2,y_2)$ then $x_1=0$ and
$x_2=2\pi$ or vice versa
\next
(ii) for each $t\in I$ the set $\phi(\{t\}\times [0,2\pi]\times T^2)$
is a hypersurface of constant mean curvature $t$.
\next
(iii) the pull-back of the metric under $\phi$ has the form
$$-\alpha^2 dt^2+A^2[(dx+\beta^1 dt)^2+
a^2\tilde g_{AB}(dy^A+\beta^A dt)(dy^B+\beta^B dt)]\eqno(2.5)$$

\noindent
The functions $\alpha$, $\beta^a$, $A$ and $\tilde g_{AB}$ depend on
$t$ and $x$ and $\tilde g_{AB}$ has unit determinant. They satisfy
$\alpha(t,2\pi)=\alpha(t,0)$, $A(t,2\pi)=A(t,0)$, 
$\beta^1(t,2\pi)=\beta^1(t,0)=0$, $\beta^B(t,0)=0$, 
$\tilde g_{AB}(t,2\pi)=n^C_An^D_B\tilde g_{CD}(t,0)$, 
where $n^A_B$ is an element of $GL(2,\Z)$. The quantity $a$ depends only 
on $t$.}

\noindent
{\it Proof} It is a standard fact that, in a non-flat spacetime satisfying
the strong energy condition, a neighbourhood of a compact CMC 
hypersurface with non-zero mean curvature can be foliated by constant mean 
curvature hypersurfaces and that the mean curvature of these hypersurfaces 
can be used as a time coordinate. If the $U(1)\times U(1)$-symmetry is
global then it follows from the uniqueness of CMC hypersurfaces that they
are unions of surfaces of symmetry. If the symmetry of the data is only local 
then some more care is needed, but since an almost identical argument has been
given in [17] the details are omitted here. The mean curvature of 
the Cauchy hypersurface in the assumption of the lemma cannot be zero. To see 
this consider the Hamiltonian constraint
$$R-k^{ab}k_{ab}+(\tr k)^2=16\pi\rho\eqno(2.6)$$
When the mean curvature is zero this implies that the scalar curvature
$R$ is non-negative. The topology of the Cauchy hypersurface is such that
any metric with non-negative scalar curvature must be flat. For its
universal cover is diffeomorphic to $\R^3$. This implies ([14], p. 324)
that the Cauchy hypersurface admits no metric of positive scalar curvature and
it is well-known that a compact 3-manifold satisfying the latter condition 
admits no non-flat metrics of non-negative scalar curvature. Hence the 
induced metric on the Cauchy hypersurface is flat and, from the Hamiltonian
constraint the second fundamental form and the energy density are zero.
It follows from the dominant energy condition that the spacetime is
vacuum everywhere and uniqueness in the Cauchy problem for the
vacuum Einstein equations shows that the spacetime is flat. Since 
the spacetime is by hypothesis non-flat, it follows that a maximal 
hypersurface is impossible. Let $U$ be an open neighbourhood of the
Cauchy hypersurface covered by a CMC foliation and let $t$ be the function
on $U$ which is equal to the mean curvature of the leaf of the
foliation on which the point lies. Choose a surface of symmetry
$F_0$ in the initial hypersurface $t=$const. and identify this
with surfaces in the other hypersurfaces $t=$const. by means of 
geodesics which start on $F_0$ orthogonal to the Cauchy hypersurface. 
Construct a mapping on each hypersurface $t=$const. 
from $[0,2\pi]\times T^2$ in the way described above. Putting 
together these mappings for all values of $t$ occurring in the
foliation of $U$ gives the mapping whose existence is asserted by the 
lemma.
\halmos

\vskip 20pt\noindent
{\bf 3. Estimates for the Hawking mass and area radius}

In this section certain general estimates for spacetimes with local
$U(1)\times U(1)$ symmetry are derived. A solution of the
Einstein constraint equations consists of a 3-dimensional manifold
$S$ and a Riemannian metric $h_{ab}$, a symmetric tensor $k_{ab}$,
a real-valued function $\rho$ and a covector $j_a$ on $M$ which
satisfy the Hamiltonian constraint (2.6) and the momentum constraint
$$\nabla^a k_{ab}-\nabla_b(\tr k)=8\pi j_b\eqno(3.1)$$
In this paper it is always assumed that the dominant energy condition 
holds and this implies that $\rho\ge |j_a|$. Suppose now that $S$ is covered
by a Gaussian foliation. In other words, if $F_0$ is a fixed leaf of the
foliation, any other leaf is obtained by going a fixed distance along the
geodesics which start normal to $F_0$. If we think of $S$ as being 
embedded in spacetime, then the resulting embedding of each leaf $F$
in this spacetime defines various geometrical objects on $F$, as is
always the case for an embedding of pseudo-Riemannian manifolds.
We present the definition of these objects in terms of such
an embedding, but in fact they are uniquely defined by the initial
data. There is a preferred orthonormal basis of the normal bundle of $F$ 
in spacetime, where the first vector is normal to $S$ and the second
vector tangent to $S$. These vectors are defined uniquely up to sign
by this condition. The geometric objects defined on $F$ by the
embedding are then the induced metric, a second fundamental form
associated to each normal vector and a 1-form, which is the
representation of the normal connection in the given normal basis.
The two second fundamental forms will be denoted by $\kappa_{AB}$
and $\lambda_{AB}$ respectively and the 1-form representing the
normal connection will be denoted by $\eta_A$. (Here upper case
Roman indices are used for objects intrinsic to $F$. Indices of objects
of this kind are raised and lowered using the induced metric $g_{AB}$ and
its inverse.) In a vacuum spacetime
the freedom in $\eta_A$ consists of just two spacetime constants. These
are the twist constants referred to in the introduction, whose vanishing
is one of the defining conditions of Gowdy spacetimes. If $u$
is an arc length parameter along the normal geodesics, the constraints
can be written in the following form:
$$\eqalignno{
&\d_u(\tr\lambda+\tr\kappa)=H(\tr\lambda+\tr\kappa)+\nabla^A\eta_A+K&  \cr
\qquad &-8\pi(\rho+J)-\f34(\tr\lambda+\tr\kappa)^2-\f12(\tilde\lambda^{AB}+
\tilde\kappa^{AB})(\tilde\lambda_{AB}+\tilde\kappa_{AB})-\eta^A\eta_A 
&(3.2) \cr
&\d_u(\tr\lambda-\tr\kappa)=-H(\tr\lambda-\tr\kappa)-\nabla^A\eta_A+K&  \cr
\qquad &-8\pi(\rho-J)-\f34(\tr\lambda-\tr\kappa)^2-\f12(\tilde\lambda^{AB}-
\tilde\kappa^{AB})(\tilde\lambda_{AB}-\tilde\kappa_{AB})-\eta^A\eta_A 
&(3.3)\cr
&\d_u\eta_A=-(\tr\lambda)\eta_A-\nabla^B\kappa_{AB}-8\pi j_A&(3.4)}
$$
Here $H$ is the trace of the second fundamental form $k_{ab}$ (i.e. the
mean curvature), $K$ is the Gaussian curvature of $F$, $\tilde\kappa_{AB}$
and $\tilde\lambda_{AB}$ are the trace-free parts of $\kappa_{AB}$ and
$\lambda_{AB}$ respectively and $J$ is the contraction of the unit
normal vector to $F$ in $S$ with $j_a$. This way of writing the 
constraints generalizes an approach used by Malec and \'O Murchadha[15],
for spherically symmetric asymptotically flat spacetimes, by the
author[17] for spatially compact surface symmetric spacetimes and 
by Chru\'sciel[5] for vacuum spacetimes with $U(1)\times U(1)$
symmetry. In the following these equations are only used in
the case of local $U(1)\times U(1)$ symmetry. It should, however,
be noted that the form of the equations suggests that there may
exist an analogue in the general case. The terms which cause 
difficulties in general are those containing derivatives 
tangential to the foliation by surfaces $F$. These are of the
form $\nabla^A\eta_A$ and $K$. The integral of the first of
these over $F$ is zero, while the integral of the second is a
constant only depending on the topology as a consequence of the
Gauss-Bonnet theorem. Thus integrating over $F$ eliminates the
tangential derivatives from equations (3.2) and (3.3).

Consider now the case of local $U(1)\times U(1)$ symmetry, with
the Gaussian foliation being that by surfaces of symmetry. Let
$\theta=\tr\lambda-\tr\kappa$, $\theta'=\tr\lambda+\tr\kappa$.
These are the expansions of the two families of null geodesics
orthogonal to $F$. Let the area radius $r$ be the square root
of the area of $F$. The Hawking mass is defined by
$m=-\f12 r\nabla^\alpha r\nabla_\alpha r$. In this case
$\nabla^A\eta_A=K=0$ and equations (3.2) and (3.3) become:
$$\eqalignno{
\d_u\theta&=-H\theta-P&(3.5)  \cr
\d_u\theta'&=H\theta'-P'&(3.6)}$$
where the quantities $P$ and $P'$ are non-negative. It is also
useful, following [15], to write these equations in the alternative
form:
$$\eqalign{
&\d_u(r\theta)=-Q-{1\over 4r}(\theta^2 r^2
+4\theta Hr^2+\theta r(\theta r-\theta' r))              \cr
&\d_u(r\theta')=-Q'-{1\over 4r}(\theta^{'2} r^2
-4\theta' Hr^2+\theta'r(\theta'r-\theta r))}\eqno(3.7)$$
where the quantities $Q$ and $Q'$ are non-negative. Consider now
a symmetric Cauchy hypersurface $S$, i.e. one which is a union of surfaces
of symmetry. Denote the maximum value attained by $r\theta$ and 
$r\theta'$ on this hypersurface by $M_+$ and the minimum by $M_-$.
Let $x_0$ be a point where the maximum is attained and suppose without
loss of generality that $\theta(x_0)\ge\theta'(x_0)$. Since $x_0$
is a critical point of $r\theta$, it follows from (3.7) that at that
point either $r\theta\le 0$ or
$$\theta^2 r^2+(4Hr)(\theta r)\le 0\eqno(3.8)$$
It follows that $M_+\le 4|Hr|$. Similarly, $M_-\ge -4|Hr|$. These
inequalities show that $\theta$ and $\theta'$ can be bounded in modulus by
$4|H|$. The Hawking mass is related to the area radius and
the expansions by $-2m/r=\f14r^2\theta\theta'$. Thus in a spacetime with local
$U(1)\times U(1)$ symmetry which is foliated by compact CMC hypersurfaces with
the mean curvature varying in a finite interval $(t_1,t_2)$ and which 
satisfies the dominant energy condition, if $r$ is bounded then $2m/r$ is 
bounded. These equations can also be used to prove a kind of positive mass 
theorem.

\noindent
{\bf Proposition 3.1} {\it Let $(M,g)$ be a spatially compact spacetime 
with local $U(1)\times U(1)$ symmetry which satisfies the 
dominant energy condition. Then the Hawking mass of
each surface of symmetry is non-negative and if the Hawking 
mass of any surface of symmetry is zero the spacetime is flat.}

\noindent
{\it Proof} The proof is similar to that of Lemma 2.4
of [17]. If $m$ vanishes on some surface $F$ then $\theta$ or
$\theta'$ is zero there. Suppose without loss of generality that
it is $\theta$. Then it can be concluded as in the proof of Lemma
2.4 of [17] that $\theta$ and $P$ vanish on any symmetric
compact Cauchy hypersurface containing $F$. When $\theta$ is zero 
the other expansion $\theta'$ is given by the rate of change of
$r$ along the compact Cauchy hypersurface. Hence $\theta'$ must vanish
somewhere and, repeating the previous argument, $\theta'$ and
$P'$ vanish on the whole Cauchy hypersurface. Looking at the explicit
forms of $P$ and $P'$ shows that $\rho=0$, $\kappa_{AB}=0$,
$\lambda_{AB}=0$ and $\eta_A=0$ on the Cauchy hypersurface. 
The vanishing of $\lambda_{AB}$ implies that $\tilde g_{AB}$ is
independent of $x$. It follows that a linear transformation with
constant coefficients of the coordinates $y^A$ can be used to reduce
the metric $\tilde g_{AB}$ on a given Cauchy hypersurface to the form
$\delta_{AB}$. This, together with the vanishing of $\kappa_{AB}$
and $\eta_A$, shows that the initial data are plane symmetric.
The subset of spacetime where $m=0$ is closed. Because of the possibility
of deforming spacelike hypersurfaces, it is also open and must be the
whole spacetime. Hence the spacetime is plane symmetric and applying Lemma 
2.4 of [17] shows that it is flat. If the spacetime is not flat then it
follows that $\theta$ and $\theta'$ can never vanish. If they had
opposite signs then this would mean that the gradient of $r$ was
everywhere spacelike and hence that the restriction of $r$ to a
Cauchy hypersurface was strictly monotonic. This is clearly impossible,
since this restriction must have a critical point somewhere. Hence
$\theta$ and $\theta'$ have opposite signs, the gradient of $r$ 
is timelike and the Hawking mass is positive.
\halmos

\vskip 10pt\noindent
The timelike vector $\nabla_a r$ is past-pointing. For otherwise $\theta$
would be negative and $\theta'$ positive. Integrating (3.5) from $0$ to
$2\pi$ on a hypersurface of constant time would then imply that $H$ was
positive somewhere, contrary to what has already been assumed. It follows
that $r$ is non-decreasing to the future along any timelike curve and that
its value at any point with time coordinate $t_1$ is bounded by above by its
value on the hypersurface $t=t_2$ if $t_1<t_2$.

\vskip 10pt
Let $n^a$ denote the unit normal to the surfaces of symmetry in the 
hypersurfaces $t=$const. and define $K=k_{ab}n^a n^b$. Then, with respect 
to the coordinates introduced in Lemma 2.1, some of the field equations take 
the following explicit forms:
$$\eqalignno{
&\d_x^2(A^{1/2})=-\f18A^{5/2}[\f32(K-\f13t)^2-\f23t^2
+2\eta_A\eta^A+\tilde\kappa^{AB}\tilde\kappa_{AB}+\tilde\lambda^{AB}
\tilde\lambda_{AB}
+16\pi\rho]&(3.9)                                         \cr
&\d_x^2\alpha+A^{-1}\d_x A\d_x \alpha=\alpha A^2[\f32(K-\f13t)^2+\f13 t^2
&\cr
&\qquad+2\eta_A\eta^A+\tilde\kappa_{AB}\tilde\kappa^{AB}
+4\pi(\rho+\tr S)]-A^2&(3.10)                              \cr
&\d_x K+3A^{-1}\d_x AK-A^{-1}\d_x At-\tilde\kappa^{AB}
\tilde\lambda_{AB}=8\pi JA&(3.11)    \cr
&\d_x\beta^1=-a^{-1}\d_t a+\f12\alpha(3K-t)&(3.12)       \cr
&\d_t a=a[-\d_x\beta^1+\f12\alpha(3K-t)]&(3.13)          \cr
&\d_t A=-\alpha KA+\d_x(\beta^1 A)&(3.14)}               
$$ 
These equations have been written in a form which makes as clear as
possible how they differ from the form they take in the special case of
plane symmetric spacetimes. The differences are not very great and in
particular equations (3.12)-(3.14) are identical to the corresponding
equations (2.6)-(2.8) in [17]. In terms of these variables the expansions
are given by $\theta=2A^{-2}\d_x A-t+K$, $\theta'=2A^{-2}\d_x A+t-K$. 

On an interval where $H$ is bounded the quantities $\theta$ and $\theta'$
and $K=t+\f12(\theta-\theta')$ are bounded. On the other hand, It follows 
from the lapse equation (3.10) that $\alpha\le 3/t^2$. 
Integrating equation (3.13) in space shows that on a finite
time interval $a$ and $a^{-1}$ are bounded. Putting this back into the
integrated equation shows that $\d_t a$ is bounded. Equation (3.13) then
implies that $\d_x\beta^1$ is bounded. Equation (3.14) can be rewritten as
$$\d_t(\log A)-\beta^1\d_x (\log A)=-\alpha K+\d_x\beta^1\eqno(3.15)$$
Together with the bounds which have just been derived this implies that 
$A$ and its inverse are bounded. Since $r=aA$ it follows immediately that
$r$ and its inverse are bounded. The inequalities just obtained serve as a 
replacement for the bound for $m^{-1}$ obtained at the corresponding point in 
the argument in [17]. The argument of [17] would apparently not work in 
the present case because $\eta_A$ makes a contribution to the equation for 
$\nabla_a m$ which has the wrong sign. The argument used here also has the 
advantage that it only requires the matter to satisfy the dominant and strong 
energy conditions and no assumption on the positivity of the pressure is 
needed. For this reason it applies to more general matter models and in 
particular to situations where an electromagnetic field is present.

The following theorem can now be proved:

\noindent
{\bf Theorem 3.1} {\it Let a solution of the Einstein equations with local
$U(1)\times U(1)$ symmetry be given and suppose that when coordinates are 
chosen which cast the metric into the form (2.5) with constant mean
curvature time slices the time coordinate takes all values in the finite 
interval $(t_1,t_2)$. Suppose further that:

\noindent
i) the dominant and strong energy conditions hold
\next
ii) $t_2<0$
\next
Then the following quantities are bounded on the interval $(t_1,t_2)$:
$$\eqalignno{
&\alpha,\d_x\alpha,A,A^{-1},\d_x A,K,\beta^1,a,a^{-1},\d_t a&(3.16)      \cr
&\d_t A,\d_x\beta^1&(3.17)}$$}

\noindent
{\it Proof} It has already been shown that $\alpha$, $A$, $A^{-1}$, $a$,
$a^{-1}$, $K$ and $\d_x\beta^1$ are bounded. The fact that $\theta$ and 
$\theta'$ are bounded implies that $A'$ is bounded. The boundedness of 
$\d_x\beta^1$ and the fact that $\beta^1$ vanishes at one point show that
$\beta^1$ is bounded. Equation (3.14) gives a bound for $\d_t A$.
Integrating equation (3.9) over the circle and using the bounds obtained
already shows that $\int_0^{2\pi}\rho$ and $\int_0^{2\pi}(2\eta_A\eta^A+
\tilde\kappa^{AB}\tilde\kappa_{AB}+\tilde\lambda^{AB}\tilde\lambda_{AB})$
are bounded. By the dominant energy condition it follows that 
$\int_0^{2\pi}j$ and $\int_0^{2\pi}\tr S$ are bounded. Finally, integrating
(3.10) starting at a point where $\d_x\alpha=0$ gives a bound for 
$\d_x\alpha$.
\halmos

\vfil\eject

\noindent
{\bf 4. Estimates for the hyperbolic and Vlasov equations}

The field equations which are used to control $W$ and $V$ are hyperbolic.
These quantities may be thought of as describing gravitational waves. The 
fact that these equations are coupled with the matter equations and are 
themselves nonlinear means intuitively that the waves interact with the 
matter and with each other. The equations will be written in terms of a 
2+2 split of the metric. Here lower case
Roman indices refer to objects which live on the quotient of spacetime by the
symmetry group. Indices of objects of this kind are raised and lowered using 
the metric $g_{ab}$ on the quotient space and its inverse.The equations are:
$$\eqalignno{
\nabla^a(r^2\nabla_a W)&=-r^2\tanh V\nabla^a W\nabla_a V-r^2(\cosh V)^{-1}
[e^{-W}T_{22}-e^WT_{33}&           \cr
&\qquad -\f12 (e^{-W}(\eta_2)^2-e^W(\eta_3)^2)]
&(4.1)\cr
\nabla^a(r^2\nabla_a V)&=r^2\cosh V\sinh V\nabla^a W\nabla_a W
-2r^2(\cosh V)^{-1}[(T_{23}-\f12 \tilde h^{AB}T_{AB}\tilde g_{23})&     \cr
&\qquad -\f12 (\eta_2\eta_3-\f12
(\tilde h^{AB}\eta_A\eta_B)\tilde g_{23})]
&(4.2)}$$
The derivation of these equations is lengthy and it proved useful for this
purpose to make use of the calculations of Kundu [13]. Let $S_W$ and $S_V$ 
denote the right hand sides of equations (4.1) and (4.2) respectively. It 
will now be shown that the modulus of each of these quantities can be bounded 
by a constant multiple of the expression
$\rho+\eta_A\eta^A+\tilde
\kappa_{AB}\tilde\kappa^{AB}+\tilde\lambda_{AB}\tilde\lambda^{AB}$. It then 
follows from what was said in the last section that under the hypotheses of 
Theorem 3.1 the $L^1$ norms of $S_W$ and $S_V$ in space are bounded by a 
constant which does not depend on time. For this purpose it is necessary to
calculate $\tilde\lambda_{AB}$ and $\tilde\kappa_{AB}$ explicitly in terms 
of $W$ and $V$.
$$\eqalignno{
\tilde\lambda_{AB}\tilde\lambda^{AB}&=\f12 A^{-2}(\cosh^2 VW_x^2+V_x^2) 
&(4.3)         \cr
\tilde\kappa_{AB}\tilde\kappa^{AB}&=\f12\alpha^{-2} [\cosh^2 V
(W_t-\beta^1W_x)^2+(V_t-\beta^1 V_x)^2]&(4.4)
}$$
This shows that the first term on the right hand side of each of the equations
$(4.1)$ and $(4.2)$
can be bounded by $\tilde\lambda_{AB}\tilde\lambda^{AB}+\tilde\kappa_{AB}
\tilde\kappa^{AB}$. To bound the other terms on the right hand side of
(4.1) and (4.2), define an orthonormal frame on each orbit by:
$$\eqalignno{
e_2&=(Aa)^{-1}(e^{-W/2}\cosh (V/2)\d/\d y^2-e^{W/2}\sinh (V/2)\d/\d y^3)
&(4.5)             \cr
e_3&=(Aa)^{-1}(-e^{-W/2}\sinh (V/2)\d/\d y^2+e^{W/2}\cosh (V/2)\d/\d y^3)
&(4.6)}$$
Then
$$\eqalignno{
e^{-W/2}\d/\d y^2&=Aa[\cosh (V/2)e_2+\sinh (V/2)e_3]&(4.7)         \cr
e^{W/2}\d/\d y^3&=Aa[\sinh (V/2)e_2+\cosh (V/2)e_3]&(4.8)}$$
The components of the covector $\eta_A$ expressed in an orthormal frame
can be bounded in terms of $\eta^A\eta_A$. Thus if the latter expression 
is bounded it follows that the components of $\eta_A$ expressed with
respect to the basis $(e^{-W/2}\d/\d y^2, e^{W/2}\d/\d y^3)$ can be bounded
by a constant multiple of $\cosh (V/2)$ or, equivalently, by a constant
multiple of $(\cosh V)^{1/2}$. This means that $e^{-W/2}\eta_2$ and $e^{W/2}
\eta_3$ can be bounded by an expression of the form $C\eta^A\eta_A
(\cosh V)^{1/2}$ for some constant $C$. This allows the expressions on the
right hand side of equations (4.1) and (4.2) containing $\eta_A$ to be 
bounded in modulus by a constant multiple of $\eta_A\eta^A$. The terms 
involving the energy-momentum tensor can be handled in a very similar way.
The dominant energy condition implies that the components of the 
energy-momentum tensor in an orthonormal frame are bounded in modulus by
$\rho$ and using (4.7) and (4.8) allows this to be translated into a bound
on the matter terms on the right hand side of equations (4.1) and (4.2)
in terms of $\rho$.

\noindent
{\bf Lemma 4.1} {\it Under the hypotheses of Theorem 3.1 the quantities
$W$, $V$, $\eta_A$, $\beta^A$, $\d_x\beta^A$ are bounded.}

\noindent
{\it Proof} Choose some $t_3$ in the interval $(t_1,t_2)$ and let $(t_4,x_4)$ 
be some point of the quotient manifold $\bar M$ parametrized by $t$ and $x$ 
with $t_4<t_3$. (The case $t_4>t_3$ is similar.) 
The equations (4.1) and (4.2) have the same characteristics. These are
the null curves of the metric defined on $\bar M$. Let $\gamma_1$ and 
$\gamma_2$ be the two characteristics passing through $(t_4,x_4)$ and let
$(t_3,x_5)$ and $(t_3,x_6)$ be the coordinates of the points where they
meet the hypersurface $t=t_3$. The left hand side of equation (4.1) has 
the form of a divergence. Applying Stokes' theorem to the triangular
region $T$ bounded by $\gamma_1$, $\gamma_2$ and the curve $t=t_3$ gives the
identity:
$$\eqalign{\int_T S_W \alpha A dt dx&=\int_{t_4}^{t_3}
(r^2 DW/Dt)(t,\gamma_1(t)) dt
               +\int_{t_4}^{t_3}(r^2 DW/Dt)(t,\gamma_2(t)) dt             \cr
&-\int_{x_5}^{x_6}r^2(W_t-\beta_1 W_x)(t_3,x) A dx}$$
and hence, after integration by parts:
$$\eqalign{(r^2W)(t_4,x_4)&=\f12[(r^2W)(t_3,x_5)+(r^2W)(t_3,x_6)]                    \cr
&-\f12\int_{t_4}^{t_3}(2r WDr/Dt )(t,\gamma_1(t))
  -\f12\int_{t_4}^{t_3}(2r WDr/Dt )(t,\gamma_2(t))                        \cr
&-\f12\int_{x_5}^{x_6}r^2(W_t-\beta_1 W_x)
(t_3,x) A dx-\f12\int_T S_W \alpha A dt dx}\eqno(4.9)$$
Here $D/Dt$ denotes a derivative in the direction of the characteristic
along which the integration is carried out, with this characteristic
being parametrized with respect to $t$. In other words, for any function
$f$, $Df/Dt=d/dt(f(\gamma(t)))$. Most of the quantities in
(4.9) are already known to be bounded. This is in particular true of
$Dr/Dt$. Thus the following inequality holds:
$$\|r^2 W(t_3-t)\|_\infty\le C\left[1+\int_0^t\left(\|r^2 W(t_3-s)\|_\infty
+\int_{\gamma_1(t_3-s)}^{\gamma_2(t_3-s)} |S_W(s,x)| dx\right) ds\right]
\eqno(4.10)$$
Since $\int_0^{2\pi} |S_W(t,x)| dx$ is known to be bounded and under the given
hypotheses the number of times the characteristics can go around the circle
between $t=t_1$ and $t=t_3$ is bounded, it follows from Gronwall's
inequality that $W$ is bounded. The same kind of argument shows that $V$ is
bounded. The remaining conclusions of the lemma are simple consequences of the
boundedness of $W$ and $V$, as will now be shown. The momentum constraint 
implies that:
$$\d_x(A^2\eta_A)=8\pi A^2j_A\eqno(4.11)$$
which means that $(A^2\eta_A)(t,x_1)-(A^2\eta_A)(t,x_2)$ is bounded 
independently of $t$, $x_1$ and $x_2$. On the other hand the 
boundedness of $\int_0^{2\pi}\eta^A\eta_A(x)dx$ together with that of $V$ 
and $W$ shows that $\int_0^{2\pi}|A^2\eta_A(x)|dx$ is bounded. These two 
facts together show that $\eta_A$ is bounded. The definition of the second 
fundamental form gives the equation:
$$\tilde g_{AB}\d_x\beta^B=2\alpha A^{-1}a^{-2}\eta_A\eqno(4.12)$$
This means that $\d_x\beta^A$ is bounded and, remembering that by definition 
$\beta^A(0)=0$, this implies that $\beta^A$ is bounded. This completes the
proof.
\halmos

Everything which has been done up to now consists of obtaining bounds for
parts of the geometry using nothing about the matter model except the
dominant energy and non-negative pressures conditions. Now the special
case of the Vlasov equation will be considered. In this class of spacetimes
the Vlasov equation for particles of unit mass takes the following form:
$$\d f/\d t+(\alpha A^{-1}(v^1/v^0)-\beta^1)\d f/\d x+F^i\d f/d v^i=0
\eqno(4.13)$$
Here the quantities $F^i$ are functions of $t$, the quantities listed in
(3.16) and (3.17), $\beta^A$ and their spatial derivatives, $\eta_A$ and
the first derivatives of $W$ and $V$ with respect to $t$ and $x$. They 
depend linearly on the derivatives of $W$ and $V$. The mass shell condition
$v^0=\sqrt{1+\delta_{ij}v^iv^j}$ defines $v^0$ in terms of $v^i$. The 
characteristics of the equation (4.13) satisfy the system:
$$\eqalign{
dx^i/ds=[\alpha A^{-1}(v^1/v^0)-\beta^1]\delta^i_1        \cr
dv^i/ds=F^i}\eqno(4.14)$$
The spacetimes considered here have two local Killing vectors. If $k^\alpha$
is a Killing vector in any spacetime and $p^\alpha$ the unit tangent vector 
to a timelike geodesic, then the quantity $p^\alpha k_\alpha$ is conserved
along the geodesic. This allows two conserved quantities for the equations
(4.14) to be derived. They can be computed, using (4.7) and (4.8) to be:
$$\eqalign{
&Aae^{W/2}[\cosh (V/2)v^2+\sinh (V/2)v^3]                \cr
&Aae^{-W/2}[\sinh (V/2)v^2+\cosh (V/2)v^3]}\eqno(4.15)$$
It is easy to solve for $v^2$ and $v^3$ in terms of these two conserved
quantities and the boundedness of $W$ and $V$ implies that $v^2$ and $v^3$
are bounded along a characteristic. Consider now a solution of the 
Einstein-Vlasov system where the initial datum for the distribution function
has compact support. Let $P(t)$ be the supremum of $|v|$ over the support
of $f(t)$. Since $a$, $A$, $W$ and $V$ have already been controlled pointwise
the components $T_{AB}$ of the energy-momentum tensor occurring on the right
hand side of (4.1) and (4.2) can be estimated in terms of the corresponding
frame components. Looking at the explicit expressions for these frame 
components and using the boundedness of $v^2$ and $v^3$ in the support of
$f$ shows that: 
$$\|T_{AB}(t)\|_\infty\le CP(t)\eqno(4.16)$$
where $C$ is a constant which only depends on the initial data. To make use of
(4.16) an estimate for $v^1$ must be obtained. Define:
$$Q(t)=\|\d_x W(t)\|_\infty+\|\d_t W(t)\|_\infty+\|\d_x V(t)\|_\infty
+\|\d_t V(t)\|_\infty\eqno(4.17)$$

\noindent
{\bf Lemma 4.2} {\it If the hypotheses of Theorem 3.1 are satisfied by a 
solution of the Einstein-Vlasov system then the following inequality holds for 
$t_4<t_3$:
$$1+P(t_4)\le C\left(1+P(t_3)+\int_0^{t_3-t_4}1+P(t_3-t)
+Q(t_3-t) dt\right)\eqno(4.18)$$
The analogous inequality holds for $t_4>t_3$.} 

\noindent
{\it Proof} In a 3+1 decomposition of a general spacetime the Vlasov  
equation takes the following form when expressed in terms of frame components:
$$\eqalign{
&\d f/\d t+(\alpha v^i/v^0 e^a_i-\beta^a)\d f/\d x^a 
\cr &\qquad
-[e_i(\alpha)v^0+\alpha(-k_{ab}e^a_ie^b_j+\gamma^i_{0j})v^j
+\alpha\gamma^i_{jk}v^jv^k/v^0]\d f/\d v^i=0}\eqno(4.19)$$
Here $\gamma^i_{0j}$ and $\gamma^i_{jk}$ are Ricci rotation coefficients.
Consider the terms appearing in $F^1$ in the case of the symmetry considered
here which contain derivatives of $V$ and $W$. No such terms arise from the 
terms in (4.19) involving the derivatives of $\alpha$ and the second 
fundamental form. To go further it is necessary to have more explicit 
expressions for the rotation coefficients. The four-dimensional rotation
coefficients $\gamma^i_{jk}$ are identical with corresponding three-dimensional
ones while:
$$\gamma^i_{0j}=-\alpha^{-1}\gamma^i_{kj}\theta^k_a\beta^a
+\f12\alpha^{-1}(e_j^a\nabla_a\beta^b\theta^i_b-\delta^{is}
e^a_s\nabla_a\beta^b\theta^t_b\delta_{jt}+c^i_j-\delta^{is}c^t_s
\delta_{jt})\eqno(4.20)$$
Here $c^i_j=e^a_j\d_t\theta^i_a$. Each term in the expressions for the
coefficients of the Vlasov equation is either independent of the derivatives
of $W$ and $V$, in which case it is bounded as a consequence of the estimates
already proved, or it is linear in these derivatives. Consider now the 
equation for $v^1$ in the characteristic system. The estimate (4.18) is
obtained by considering the dependence on $v$ of those terms which are linear
in the derivatives of $W$ and $V$. The quantity $v^i$ is bounded unless $i=1$
while the quantity $v^jv^k/v^0$ is bounded unless $j=k=1$. However, by the
symmetry properties of the rotation coefficients, $\gamma^1_{j1}=0$. Hence 
all terms on the right hand side of the equation for $v^1$ can be bounded by
an expression of the form $C(1+P+Q)$.
\halmos

\vskip 10pt\noindent
{\bf Lemma 4.3} {\it If the hypotheses of Theorem 3.1 are satisfied by a 
solution 
of the Einstein-Vlasov system then the quantities $P$, $\d_t W$, $\d_t V$
$\d_x W$, $\d_x V$, $\rho$, $\alpha^{-1}$, the derivative with respect to
$x$ of all the quantities in (3.16) and (3.17), $\d_x\eta_A$ and 
$\d_x^2 \beta^A$ are bounded.}

\noindent
{\it Proof} The first step is to obtain an estimate for the first derivatives 
of $W$ and $V$. In order to do this, it is useful to write equations (4.1)
and (4.2) in a slightly different way.
$$\eqalignno{
\nabla^a\nabla_a W+\tanh V\nabla^a W\nabla_a V&=-(2/r)\nabla^a r\nabla_a W
-(\cosh V)^{-1}[e^{-W}T_{22}-e^WT_{33}&           \cr
&\qquad -\f12 (e^{-W}(\eta_2)^2-e^W(\eta_3)^2)]&(4.21)   \cr 
\nabla^a\nabla_a V-\sinh V\cosh V\nabla^a W\nabla_a W&=-(2/r)\nabla^a r
\nabla_a V-2(\cosh V)^{-1}[(T_{23}-\f12 \tilde h^{AB}T_{AB}
\tilde g_{23})&     \cr
&\qquad -\f12 (\eta_2\eta_3-\f12
(\tilde h^{AB}\eta_A\eta_B)\tilde g_{23})]&(4.22)}$$
The advantage of this is that if the right hand sides of (4.21) and (4.22)
are replaced by zero the resulting equations are those for a wave map
(hyperbolic harmonic map) with target space $\R^2$, endowed with the
metric $\cosh^2 V dW^2+dV^2$. This is a representation of the standard metric
of the hyperbolic plane in a certain coordinate system. It is natural to try 
to generalize estimates which have been used in the study of wave maps to the 
present situation. Here this is done with an estimate of Gu [11], who 
used it to prove global existence of classical solutions in the Cauchy problem
for wave maps defined on two-dimensional Minkowski space. Define two null
vectors on the two-dimensional space coordinatized by $t$ and $r$ by
$$\eqalign{
e_+&=\alpha^{-1}(\d/\d t-\beta\d/\d x)+A^{-1}\d/\d x           \cr
e_-&=\alpha^{-1}(\d/\d t-\beta\d/\d x)-A^{-1}\d/\d x}\eqno(4.23)$$
The (2-dimensional) covariant derivatives $\nabla_{e_-} e_+$ and 
$\nabla_{e_+} e_-$ are given by:
$$\eqalign{
\nabla_{e_-} e_+=\alpha^{-1}(b_{++}e_+ +b_{+-}e_-)           \cr
\nabla_{e_+} e_-=\alpha^{-1}(b_{-+}e_- +b_{--}e_-)}\eqno(4.24)$$
for some bounded functions $b_{++}$, $b_{+-}$, $b_{-+}$ and $b_{--}$. The
normalization chosen for the vectors $e_+$ and $e_-$ here is important,
since otherwise the covariant derivatives could contain the time derivatives
of $\alpha$ and $\beta^1$, quantities which have not yet been shown to be
bounded. Let $E_+$ and $E_-$ be the images of $e_+$ and $e_-$ under the
wave map, i.e. 
$$\eqalign{
E_+=e_+(W)\d/\d W+e_+(V)\d/\d V                 \cr
E_-=e_-(W)\d/\d W+e_-(V)\d/\d V}\eqno(4.25)$$
Let $\gamma_1$ and $\gamma_2$ be integral curves of $e_-$ and $e_+$
respectively and let $\hat\gamma_i=(W\circ\gamma_i,V\circ\gamma_i)$.
The observation of Gu is that the equations obtained from (4.21) and
(4.22) by replacing the right hand side by zero and the given metric
by the flat metric just say that $E_+$ is parallelly transported along
$\hat\gamma_1$ and that $E_-$ is parallelly transported along $\hat\gamma_2$.
A similar calculation can be done for the equations (4.21) and (4.22) and
this gives rise to the following equation along $\hat\gamma_1$ (and an 
analogous equation along $\hat\gamma_2$):
$$\nabla_{\alpha E_-}E_+=(b_{++}-r^{-1}\alpha e_-(r))E_+ 
+(b_{+-}-r^{-1}\alpha e_+(r))E_- 
+B_- \eqno(4.26)$$
where $B_-$ satisfies an inequality of the form 
$|B_-|\le C(1+\|T_{AB}\|_\infty)$.
These equations allow the lengths of the vectors $E_+$ and $E_-$ to be
controlled. Multiplying (4.26) and the analogous equation 
for $E_+$ by $\alpha$ allows the following inequality to be derived:
$$Q(t_4)\le C[Q(t_3)+\int_0^{t_3-t_4}1+Q(t_3-t)+\|T_{AB}(t_3-t)\|_\infty dt]
\eqno(4.27)$$
Putting together (4.16), (4.18) and (4.27) gives:
$$(1+P+Q)(t_4)\le(1+P+Q)(t_3)+C\int_0^{t_3-t_4}(1+P+Q)(t_3-t) dt
\eqno(4.28)$$
Hence by Gronwall's lemma $P$, $\d_t W$, $\d_t V$, $\d_x W$ and $\d_x V$
are bounded. It then follows immediately that $\rho$ is bounded and (3.10)
shows that $\alpha^{-1}$ is bounded. The equations (3.9)-(3.14) can be used
directly to show that the first derivatives with respect to $x$ of all the 
quantities in (3.16) and (3.17) are bounded. It follows from (4.11) that
$\d_x\eta_A$ is bounded and from (4.12) that $\d_x^2\beta^A$ is bounded.
\halmos

\vskip 10pt\noindent
{\bf Lemma 4.4} {\it If the hypotheses of Theorem 3.1 are satisfied by a 
solution 
of the Einstein-Vlasov system then the second derivatives of $W$ and $V$ and
the first derivatives of $f$ are bounded.}

\noindent
{\it Proof} If $f$ were zero (the vacuum case) then it would be rather
simple to prove this theorem, since the equations obtained by differentiating 
the equations for $W$ and $V$ with respect to $x$ are linear in the highest 
order derivatives in that case.
With the coupling to $f$ things are less straightforward. When the Vlasov
equation is differentiated with respect to $x$ terms come up which involve
second derivatives of $W$ and $V$ multiplied by first derivatives of $f$.
In other words, there are terms which are quadratic in the quantities to
be estimated and this precludes a direct application of Gronwall's 
inequality. This problem can be solved using a device of Glassey and Strauss
[10], which can be seen in a particularly simple form, adequate for the 
present application, in the paper [8] of Glassey and Schaeffer (see also
[9]). The equation for $W$ can be written in the following form:
$$l^a\nabla_a(n^b\nabla_b W)=(Y_1(W,V)l^a+Y_2(W,V)n^a)\nabla_a W+Z(W,V)
\eqno(4.29)$$
where $Z(W,V)$ contains no derivatives of $W$ or $V$ and $Y_1(W,V)$ and 
$Y_2(W,V)$ contain them at most linearly. Here, for ease of notation, $l=e_+$ 
and $n=e_-$. The equation for $V$ can of course be written
in a similar form. There are also alternative forms of both of these 
equations where the roles of $l$ and $n$ are interchanged. Differentiating
equation (4.29) with respect to $x$ gives an equation of the form
$$l^a\nabla_a(\d_x(n^b\nabla_b W))=(Y_1(W,V)l^a+Y_2(W,V)n^a)
\d_x(\nabla^a W)+\tilde Z(W,V)\eqno(4.30)$$ 
where the expression $\tilde Z(W,V)$ does 
not depend on second derivatives of $W$ and $V$. Suppose now that we integrate
the equation (4.30) along the characteristic which is an integral curve
of $l^a$. The only terms which cannot be bounded straightforwardly (even
before integration) are those which contain derivatives of the energy
momentum tensor with respect to $x$. It will now be shown how a typical 
term of this type can be handled. The others which occur can be taken care
of in a strictly analogous way.

The term which is to be bounded is:
$$\int_0^{t_3-t_4}[(e^W\cosh V)^{-1}\d_x T_{22}](t_3-t) dt\eqno(4.31)$$
In fact the coordinate components of the energy-momentum tensor may be 
replaced by frame components at this stage since their spatial derivatives 
only differ by terms which are bounded. Substituting the definition of the
frame component $T(e_2,e_2)$ into the expression of interest gives
$$\int_0^{t_3-t_4}\int [(e^W\cosh V)^{-1} (v_2)^2(1+|v|^2)^{-1}]\d_x f dv dt
\eqno(4.32)$$
The idea of [10] is to express $\d_x$ as a linear combination of $l$ and 
the vector 
$$m=\d/\d t+(\alpha A^{-1}(v^1/v^0)-\beta^1)\d/\d x\eqno(4.33)$$
The result is:
$$\d/\d x=\alpha^{-1}A(1-v^1/v^0)^{-1}(l-m)\eqno(4.34)$$
This allows the integral in (4.32) to be rewritten as a sum of two terms,
one containing $l$ and the other containing $m$. Now it is possible to
substitute for $mf$ using the Vlasov equation and the result contains
only derivatives of $f$ with respect to the velocity variables. These
derivatives can be eliminated by an integration by parts in $v$ and the
result is a bounded quantity. The other term is equal to
$$\eqalign{
&\int\int_0^{t_3-t_4} [\alpha^{-1}A(e^W\cosh V)^{-1} 
(v_2)^2(1+|v|^2)^{-1} l^a\nabla_a f] (\gamma_1(t_3-t)) dt 
(1-v^1/v^0)^{-1} dv       \cr
&=\int [(\alpha^{-1}A(e^W\cosh V)^{-1})(t_4,x_4)f(t_4,x_4,v)\cr
&\qquad-(\alpha^{-1}A(e^W\cosh V)^{-1})(t_3,x_5) f(t_3,x_5,v)] 
(1-v^1/v^0)^{-1} dv +\ldots }\eqno(4.35)$$
Thus it is also bounded. The same trick can be applied when $W$ is replaced
by $V$ and when $l$ and $n$ are interchanged. (In the last case $\d/\d x$
must be replaced by a combination of $m$ and $n$.) The result of all this
is that if 
$$Q_1=\|\d_x (l^a\nabla_a W)\|_\infty+\|\d_x (n^a\nabla_a W)\|_\infty
+\|(l^a\nabla_a V)\|_\infty+\|(n^a\nabla_a V)\|_\infty\eqno(4.36)$$ 
then an estimate of the form:
$$1+Q_1(t_4)\le 1+Q_1(t_3)+C\int_0^{t_3-t_4} (1+Q_1(t_3-t)) dt\eqno(4.37)$$
is obtained. It follows from Gronwall's inequality that $Q_1$ is bounded.
Hence the derivatives $W_{xx}$, $W_{tx}$, $V_{xx}$ and $V_{tx}$ are
bounded. Using this information in the equations obtained by differentiating
the Vlasov equation with respect to $x$ or $v$ shows that the first 
derivatives of $f$ with respect to these variables are bounded.
\halmos

\vskip 20pt\noindent
\noindent
{\bf 5. The main result}

In this section the estimates collected in Section 4 are applied to
prove the main result. First one last auxiliary lemma is required.

\noindent
{\bf Lemma 5.1} {\it Consider a CMC initial data set for the Einstein-Vlasov 
system with local $U(1)\times U(1)$ symmetry. Then there exists a local Cauchy
evolution of this data which has local $U(1)\times U(1)$ symmetry, so that
the hypotheses of Lemma 2.1 are satisfied. Consider next a family of initial 
data sets of this type on the same manifold such that:

\noindent
(i) the data in the family are uniformly bounded in the $C^\infty$ topology 
\next
(ii) the metrics are uniformly positive definite 
\next
(iii) the supports of the distribution functions are contained in a common
compact set
\next
(iv) the mean curvatures are uniformly bounded away from zero
\next
Then the time interval in the conclusion of Lemma 2.1 can be chosen 
uniformly for the Cauchy evolutions of all data in the family.} 

\noindent
{\it Proof} The first statement of the proof is essentially a direct 
consequence of the standard local existence theorem for the Einstein-Vlasov
system and for CMC hypersurfaces and the fact that the resulting spacetimes
inherit any symmetry which is present. When there is only local symmetry
the inheritance of symmetry argument should be applied to the universal
cover (cf. [17]). The second part of the lemma, concerning families is a 
consequence of the stability of various operations. Firstly, the statement
is used that the time of existence of a solution of the Einstein-Vlasov system,
measured with respect to an appropriate time coordinate, depends only on the
size of the initial data and that on a fixed closed time interval the solution
depends continuously on the initial data. Secondly, the fact is used that the 
interval on which a CMC foliation exists in a neighbourhood of a given CMC 
hypersurface depends only on the size of the metric coefficients in 
an appropriate coordinate system and a positive lower bound for the lapse 
function, provided the mean curvature of the starting hypersurface remains 
bounded away from zero. 
\halmos

\vskip 10pt\noindent
{\bf Theorem 5.1} {\it Let $(M,g,f)$ be a $C^\infty$ solution of the 
Einstein-Vlasov system with local $U(1)\times U(1)$ symmetry which is the
maximal globally hyperbolic development of data on a hypersurface of
constant mean curvature $H_0<0$. Then the part of the spacetime to the
past of the initial hypersurface can be covered by a foliation of CMC
hypersurfaces with the mean curvature taking all values in the interval
$(-\infty,H_0]$. Moreover, the CMC foliation can be extended to the 
future of the initial hypersurface in such a way that the mean curvature
attains all negative real values.}

\noindent
{\it Proof} Let $T$ be the largest number (possibly infinite) such that the 
local foliation by CMC hypersurfaces which exists near the initial 
hypersurface can be extended so that the mean curvature takes on all values
in the interval $(-T,H_0)$. Suppose that $T$ is finite. Then Theorem 3.1
and the results of Section 4 imply the boundedness of many quantities on
the interval $(-T,H_0]$. It will now be shown by induction that the spatial 
derivatives of all orders of all quantities of interest are bounded on the
given interval. the inductive hypothesis is that the following quantities are 
bounded:
$$\eqalign{
&D^n f, D^{n+1} W, D^n (\d_t W), D^{n+1} V, D^n (\d_t V),
D^{n+1}\alpha,                  \cr
&D^{n+1}\beta^1, D^{n+1} A, D^n(\d_t A), D^n K,
D^n\eta_A, D^{n+1}\beta^A}\eqno(5.1)$$  
It follows from the results of Section 4, and in particular Lemma 4.4, that 
the inductive hypothesis is satisfied for $n=1$. Suppose now that it is
satisfied for a given value of $n$. Then it follows immediately from the
field equations (3.9)-(3.14) and (4.11)-(4.12) that all quantities which are 
required to be bounded by the inductive hypothesis at the next step are 
bounded, except possibly for the relevant derivatives of $f$, $W$ and $V$. 
Consider the equation obtained by differentiating the Vlasov equation $n+1$ 
times with respect to $x$. There results a linear equation for $D^{n+1} f$ 
with coefficients which are known to be bounded, except for terms involving 
derivatives of $W$ and $V$ in the inhomogeneous term. If 
$F_n(t)=\|D^n f\|_\infty$ and
$$Q_n(t)=\|D^{n+1}W\|_\infty+\|D^{n+1}V\|_\infty+\|D^n (\d_t W)\|_\infty+
\|D^n (\d_t V)\|_\infty$$
then this equation implies an inequality of the form
$$F_{n+1}(t)\le F_{n+1}(t_0)+C\int_0^{t_0-t} F_{n+1}(t_0-s)
+Q_{n+1}(t_0-s) ds\eqno(5.2)$$
Similarly, differentiating equations (4.29) and (4.30) $n$ times with
respect to $x$ gives a linear system of equations for derivatives of $V$
and $W$ with coefficients which are known to be bounded, except for terms
involving derivatives of order $n+1$ of matter quantities in the 
inhomogeneous term. Hence:
$$G_{n+1}(t)\le G_{n+1}(0)+C\int_0^{t_0-t} F_{n+1}(t_0-s)
+G_{n+1}(t_0-s) ds\eqno(5.3)$$
Putting together (5.2) and (5.3) and applying Gronwall's inequality proves
that $F_{n+1}$ and $G_{n+1}$ are bounded and completes the inductive step. 
Thus the quantities in (5.1) are bounded for all $n$. Consider now the data 
obtained by restricting the given solution to the hypersurfaces $t=$const. By 
what has just been proved, this family of data satisfies the conditions of
Lemma 5.1. Hence there exists some $\epsilon>0$ such that each of these
initial data has a corresponding solution on a time interval of length 
$\epsilon$. Hence the original solution extends to the interval $(-T-\epsilon,
H_0)$, contradicting the maximality of $T$. It follows that in fact $T=\infty$,
as desired. This means in particular that the spacetime has a crushing
singularity in the past, and hence that the CMC foliation covers the entire 
past of the initial hypersurface. 

Now let $T'$ be the largest number such that the CMC foliation can be extended
to the interval $(-\infty,T')$. Since the spacetime contains no compact
maximal hypersurface $T'\le 0$. If $T'$ were strictly less than zero it 
could be argued as in the first part of the proof that the CMC foliation
could be extended further, which would contradict the definition of $T'$.
Hence in fact $T'=0$.
\halmos

\vskip 10pt\noindent
This argument does not prove that the entire future of the initial 
hypersurface is covered by the CMC foliation. In connection with this it is
interesting to note that if instead of assuming, as is done in this paper,
that the cosmological constant $\Lambda$ vanishes, it is assumed that 
$\Lambda<0$ then the same types of arguments apply to give a stronger
theorem. (The choice of sign convention for the cosmological constant
used here is such that $\Lambda<0$ corresponds to anti-de Sitter space.)
With $\Lambda<0$ the result is that the whole spacetime can be covered by
a CMC foliation with the mean curvature taking on all real values. The
reason for this difference can be traced to the estimate for $\alpha$
following from the lapse equation, which in general reads $\alpha\le
(\f13 t^2-\Lambda)^{-1}$.

\vskip 20pt\noindent
\noindent
{\bf 6. The case of wave maps}

In this section we consider what happens when the collisionless matter
described by the Vlasov equation is replaced by a wave map as source in
the Einstein equations. This is quite natural, given that, as was seen 
in Section 4, a wave map comes up automatically in the case of vacuum 
spacetimes. Let $(N,h)$ be a complete Riemannian manifold. If 
$(M,g)$ is a Lorentz manifold a wave map $\phi$ from $M$ to $N$ is a map
which satisfies the equation whose expression in local coordinates 
$x^\alpha$ on $M$ and $y^I$ on $N$ is:
$$\nabla_\alpha\nabla^\alpha\phi^I+\Gamma^I_{JK}\nabla_\alpha\phi^J
\nabla^\alpha\phi^K=0\eqno(6.1)$$
(Wave maps are also known as (hyperbolic) harmonic maps or nonlinear
sigma models.) The global Cauchy problem for wave maps on two-dimensional 
Minkowski space was solved by Gu[11] and for wave maps on three-dimensional 
Minkowski space which are invariant or equivariant under rotations
by Christodoulou, Shatah and Tahvildar-Zadeh [3, 4, 19]. The results
of [4] were applied to the Einstein-Maxwell equations in [2].
Associated to a wave map $\phi$ is the energy-momentum tensor:
$$T_{\alpha\beta}=[\nabla_\alpha\phi^I\nabla_\beta\phi^J-\f12(
\nabla_\gamma\phi^I\nabla^\gamma\phi^J)g_{\alpha\beta}]h_{IJ}\eqno(6.2)$$
and this can be used to couple the wave map to the Einstein equations.
In harmonic coordinates the coupled equations form a system of nonlinear
wave equations and so a local existence and uniqueness theorem can be
proved by the usual methods. The energy-momentum tensor of a wave 
map satisfies both the dominant and strong energy conditions. This
can be seen by noting that both these conditions are purely algebraic in 
nature and can be checked using normal coordinates based at a given
point of $N$. Then the energy-momentum is reduced at a point to a sum of 
terms, each of which is the energy-momentum tensor of a massless scalar
field. 

Consider now the case of a solution of the Einstein equations with local
$U(1)\times U(1)$ symmetry coupled to an invariant wave map. To say that
the wave map is invariant means that each surface of symmetry is mapped
to a single point of $N$. Since the relevant energy conditions hold,
it follows that the analogues of the results obtained in Sections
2 and 3 for the Einstein-Vlasov system are also valid for the 
Einstein-wave map system. Given that in proving Theorem 5.1 a wave map
was already estimated, albeit for a special target manifold $(N,h)$, it
appears straightforward to generalize that theorem to the case of the
Einstein-wave map system. In fact the equation of motion for the wave
map does not involve $W$, $V$ or $\eta_A$ while the combinations of matter
terms occurring in (4.1) and (4.2) vanish identically for the energy-momentum
tensor of an invariant wave map. Thus there is no direct coupling between
the wave map describing the matter and the wave-map-like equation satisfied
by $W$ and $V$. The one difficulty which occurs is
that, in contrast to the special case of the hyperbolic plane,
there is no global coordinate system on $N$ in the general case. The 
equation for an invariant wave map can be written in the form:
$$\nabla_a(r^2 \nabla^a\phi^I)+r^2\Gamma^I_{JK}\nabla_a\phi^J\nabla^a
\phi^K=0\eqno(6.3)$$
This bears a strong resemblance to equations (4.1)-(4.2), with the 
difference that there are no terms involving $\eta$ or matter quantities
in (6.3). This makes the analogue of the calculation (4.9) for the
wave map superfluous. This is just as well, since it seems difficult
to formulate an analogue of (4.9) in the case that there is no global
coordinate system on $N$. What can be done instead is to go directly
to the analogue of (4.26) for the wave map. Define:   
$$\eqalign{
\tilde E_+=e_+(\phi^I)\d/\d\phi^I                 \cr
\tilde E_-=e_-(\phi^I)\d/\d\phi^I}                \eqno(6.4)$$
Then $\tilde E_+$ and $\tilde E_-$ satisfy propagation equations like (4.26)
along $\hat\gamma_1$ and $\hat\gamma_2$ respectively. There is no term 
corresponding to $B_-$ in this case. It follows that under the hypotheses of 
Theorem 3.1 the length of the vectors $\tilde E_+$ and $\tilde E_-$ is bounded
on the given time interval. This implies a bound on the distance of any point
of the image 
under $\phi$ of this time interval from the image of the initial hypersurface.
In particular, the image of this time interval under $\phi$ is contained
in a compact subset of $N$. This compact set can be covered by a finite
number of charts, each of which can be chosen to be defined on a domain 
with compact closure in a larger chart domain. In each of these charts the
quantities $\phi^I$, $\d_t\phi^I$ and $\d_x\phi^I$ are bounded. Moreover,
any of these charts the following analogue of (4.30) holds: 
$$l^a\nabla_a(\d_x(n^b\nabla_b \phi^I))=(Y_1(\phi^I)l^a
+Y_2(\phi^I)n^a)\d_x(\nabla^a \phi^I)+\tilde Z(\phi^I)\eqno(6.5)$$
This equation and the equations obtained by differentiating it repeatedly
with respect to $x$ can be used to inductively bound all spatial derivatives 
of $\phi^I$. This proceeds essentially as in the proof of Theorem 5.1;
it is merely necessary to be careful about the different charts which occur.
In the case of a wave map define $F_n(t)$ to be the maximum over the 
finite set of charts chosen of $\|D^{n+1}\phi^I\|_\infty$ and 
$\|D^n (\d_t\phi^I)\|_\infty$. When the derivatives of lower orders are 
known to be bounded, this is equivalent to choosing for each point one chart 
which contains its image and only taking the supremum over those values.
When the quantity $F_n(t)$ is bounded the derivatives
of order $n$ of the frame components of the energy-momentum tensor are
bounded. In order to get an inequality which can be used to control
$F_n(t)$, we would like to integrate a derivative of (6.5) along a
characteristic (integral curve of $l$ or $n$). The image of this 
characteristic under $\phi$ need not be contained in a single chart.
Consider such a characteristic $\gamma$, parametrized by $t$ from $t=0$ to
$t=T$. For each $t\in [0,T]$ there exists an interval $I$, open in $[0,T]$, 
whose image under $\phi\circ\gamma$ is contained in one of the chosen charts 
on $N$. By compactness of $[0,T]$, finitely many of these intervals cover it.
It follows that there is a finite sequence of times $\{0=t_1,t_2,\ldots,t_k\}
=T$ such that $\phi\circ\gamma([t_i,t_{i+1}])$ is contained in one of the 
chosen charts for all $i$ between $1$ and $k-1$. It follows from (6.5) that
$$F_n(t_k)\le F_n(t_{k-1})e^{C(t_k-t_{k-1})}\eqno(6.6)$$
This is enough to allow $F_n$ to be bounded for all $t\in [0,T]$. Thus the
following analogue of Theorem 5.1 is obtained:

\vskip 10pt\noindent
{\bf Theorem 6.1} {\it Let $(N,h)$ be a complete Riemannian manifold and let
$(M,g,\phi)$ be a $C^\infty$ solution of the Einstein equations with local 
$U(1)\times U(1)$ symmetry coupled to an invariant wave map with target
space $(N,h)$ which is the maximal globally hyperbolic development of data on 
a hypersurface of constant mean curvature $H_0<0$. Then the part of the 
spacetime to the past of the initial hypersurface can be covered by a 
foliation of CMC hypersurfaces with the mean curvature taking all values in 
the interval $(-\infty,H_0]$. Moreover, the CMC foliation can be extended to 
the future of the initial hypersurface in such a way that the mean curvature
attains all negative real values.}

\vskip 10pt\noindent
{\it Acknowledgement} 

\noindent
I am grateful to Lars Andersson for helpful discussions.

\vskip 10pt\noindent
{\bf References}

\noindent
[1] Bartnik, R.: Remarks on cosmological spacetimes and constant mean
curvature hypersurfaces. Commun. Math. Phys. {\bf 117}, 615-624 (1988).
\next
[2] Berger, B. K., Chru\'sciel, P. T. and Moncrief, V.: On 
\lq asymptotically flat\rq\ spacetimes with $G_2$-invariant Cauchy 
surfaces. Ann. Phys. {\bf 237}, 322-354 (1995).
\next
[3] Christodoulou, D., Tahvildar-Zadeh, S.: On the regularity of 
spherically symmetric wave maps. Commun. Pure Appl. Math. {\bf 46}, 1041-1091
(1993).
\next
[4] Christodoulou, D., Tahvildar-Zadeh, S.: On the asymptotic behaviour
of spherically symmetric wave maps. Duke Math. J. {\bf 71}, 31-69 (1993).
\next
[5] Chru\'sciel, P. T.: On spacetimes with $U(1)\times U(1)$ symmetric
compact Cauchy surfaces. Ann. Phys. {\bf 202}, 100-150 (1990).
\next
[6] Chru\'sciel, P. T., Isenberg, J. and Moncrief, V.: Strong cosmic
censorship in polarised Gowdy spacetimes. Class. Quantum Grav. {\bf 7}, 
1671-1680 (1990).
\next
[7] Eardley, D., Smarr, L.: Time functions in numerical relativity: 
marginally bound dust collapse. Phys. Rev. D {\bf19}, 2239-2259 (1979).
\next
[8] Glassey, R., Schaeffer, J.: On the \lq one and one-half dimensional\rq\
relativistic Vlasov-Maxwell system. Math. Meth. Appl. Sci. {\bf 13}, 169-179 
(1990).
\next
[9] Glassey, R., Schaeffer, J.: The relativistic Vlasov-Maxwell equations
in low dimension. In: Murthy, M. K. V., Spagnolo, S. (eds.): Nonlinear 
hyperbolic equations and field theory. Pitman, London, 1992.
\next
[10] Glassey, R., Strauss, W.: Singularity formation in a collisionless
plasma could only occur at high velocities. Arch. Rat. Mech. Anal. {\bf 92},
56-90 (1986).
\next
[11] Gu, C.-H.: On the Cauchy problem for harmonic maps defined on
two-dimensional Minkowski space. Commun. Pure Appl. Math. {\bf 33}, 727-737
(1980).
\next
[12] Isenberg, J., Moncrief, V.: The existence of constant mean curvature
foliations of Gowdy 3-torus spacetimes. Commun. Math. Phys. {\bf 86}, 485-493 
(1983).
\next
[13] Kundu, P.: Projection tensor formalism for stationary axisymmetric
gravitational fields. Phys. Rev. D {\bf 18}, 4471-4479 (1978).
\next
[14] Lawson, H. B., Michelson, M.-L.: Spin geometry. Princeton University
Press, Princeton, 1989.
\next
[15] Malec, E., \'O Murchadha, N.: Optical scalars and singularity
avoidance in spherical spacetimes. Phys. Rev. D {\bf 50}, 6033-6036 (1994).
\next
[16] Marsden, J. E., Tipler, F. J.: Maximal hypersurfaces and foliations of 
constant mean curvature in general relativity. Phys. Rep. {\bf 66}, 109-139 
(1980).
\next
[17] Rendall, A. D.: Crushing singularities in spacetimes with spherical, 
plane and hyperbolic symmetry. Class. Quantum Grav. {\bf 12}, 1517-1533 (1995).
\next
[18] Scott, P.: The geometries of 3-manifolds. Bull. London Math.
Soc. {\bf 15}, 401-487 (1983). 
\next
[19] Shatah, J., Tahvildar-Zadeh, S.: Regularity of harmonic maps from the 
Minkowski space into rotationally symmetric manifolds. Commun. Pure Appl.
Math. {\bf 45}, 947-971 (1992).

\end